\documentclass[aps,floatfix]{revtex4}
\usepackage{graphics,epsfig}
\usepackage{graphicx}

\begin{document}

\title{Five-dimensional warped product space-time with time-dependent warp factor and cosmology of the four-dimensional universe}

\author{Sarbari Guha$^1$ and Subenoy Chakraborty$^2$}
\address{$^1$ Department of Physics, St. Xavier's College (Autonomous), Kolkata 700 016, India}
\address{$^2$ Department of Mathematics, Jadavpur University, Kolkata 700032, India}

\begin{abstract}
In this paper, we have studied a 5-dimensional warped product space-time with a time-dependent warp factor. This warp factor plays an important role in localizing matter to the 4-dimensional hypersurface constituting the observed universe and leads to a geometric interpretation of dynamical dark energy. The five-dimensional field equations are constructed and its solutions are obtained. The nature of modifications produced by this warp factor in the bulk geometry is discussed. The hypersurface is described by a flat FRW-type metric in the ordinary spatial dimension. It is found that the effective cosmological constant of the four-dimensional universe is a variable quantity monitored by the time-dependent warp factor. The universe is initially decelerated, but subsequently makes a transition to an accelerated phase at later times.
\end{abstract}


\maketitle

\section{Introduction}
\label{intro}
Extra dimensional theories have aroused the interest of physicists ever since the works of Kaluza and Klein \cite{kk}. These theories continue to attract wide attention owing to their success in solving some outstanding problems of physics \cite{aadd,add,khoury,rbvspv}. Several works are available on the theories of non-compact extra dimensions \cite{noncompact,Gogberashvili}. All these finally led to the 'braneworld scenario', where the ordinary standard model matter and non-gravitational fields are confined to the 4-dimensional universe, embedded in a $(4 + n)$-dimensional 'bulk' (n being the number of extra dimensions). At low energies, gravity is confined to the hypersurface along with particles, but at high energies gravity "leaks" into the higher-dimensional bulk. The change incorporated in the short distance behavior of gravity at distances much larger than the Planck length \cite{aadd,add,khoury,others,HW}, by lowering the fundamental scale of gravity through the introduction of large extra dimensions, led to extensive studies in the cosmological scenarios for the early universe \cite{Lukas,cosmo1,cosmo2,other1,other2}.

\bigskip
Among the several higher-dimensional models developed over the years, the warped braneworld model of Randall and Sundrum \cite{rs1,rs2}, with a single extra dimension, has become much popular. In their model, matter fields were localized on a 4-dimensional hypersurface in a constant curvature five-dimensional bulk furnished with mirror symmetry, with an exponential warp factor and a non-factorizable metric, even when the fifth dimension was infinite. The field equations on the corresponding 4-dimensional universe were modified by the effect of the extra dimension. Generalizations of the RS-scheme were investigated in a number of papers (see for example \cite{LR,ADDK,otherRS}). In the RS scenario, the warp factor was a function of the extra-dimensional coordinate and the metric coefficient for the extra-dimensional coordinate was constant. However, both of these parameters can be functions of both time and the extra-dimensional coordinate, with the solutions becoming much more complicated. Nevertheless, the study of five-dimensional warped product space-times and their solutions are very important for understanding the evolution of our universe.

\bigskip
Although a vast amount of literature is available on the studies of bulk spacetimes with the extra-dimensional scale factor either as a constant or a function of time or the extra-dimensional coordinate or both (see for example \cite{bdl1,bdl2,Nihei}), the effect of a time-dependent warp factor has not been studied \cite{GW,LS,KKOP,CGRT,CGKT,CGS}. In some cases (as in \cite{Neupane1}), the role of bulk matter fields have been ignored. We know that in the RS models, the exponential warp factor reflects the confining role of the negative bulk cosmological constant \cite{lrr} to localize gravity at the 4-dimensional hypersurface through the curvature of the bulk. However, the process of localization of gravity may also include some time-dependence during a particular stage of evolution. This time-dependent process of localization may be related to a time-dependent warp factor, which in turn may be related to a time-varying bulk cosmological constant. Our objective, in this paper is to study the effect of the time-dependent warp factor in the context of classical general relativity and to determine the nature of modifications produced by this warp factor in the bulk geometry as well as its consequences on the four-dimensional universe. It is found that the effective cosmological constant of the four-dimensional universe is a variable quantity monitored by the time-dependent warp factor. This leads to a geometric interpretation of dynamical dark energy. To develop the necessary theory, we have considered the RS model with the extra dimensional scale factor independent of time and a non-empty bulk, thereby ensuring the recovery of standard FRW cosmology \cite{KKOP} in the effective 4D theory at energies below the weak scale. The hypersurface is chosen to be a flat FRW-type metric in the ordinary spatial dimension.

\bigskip
The plan of the paper is as follows: In section II, the basic theoretical framework has been discussed. This has been followed by the construction of the field equations and analysis of the energy conditions in the bulk, in the next section. In section IV, the solutions to the field equations have been determined for specific cases, namely when the Weyl tensor of the bulk space-time vanishes and when the pressure in the bulk is isotropic. The effect of the warp factor on the bulk geometry as well as on the energy conditions in the bulk has been analyzed. In section V, the cosmology of the effective four-dimensional part has been analyzed. The effect of the time-dependent warp factor in modifying the usual dynamics of the gravitational field compared to that predicted from Einstein's theory, has been discussed. The vacuum energy of the corresponding hypersurface is found to be a time-varying entity monitored by the warp factor. The status of the energy conditions and the nature of the expansion of the 4-dimensional universe has been studied. For a specific solution corresponding to an isotropic bulk, the universe is initially decelerated, but subsequently makes a transition to an accelerated phase at later times. The summary of the entire exercise has been presented in Section VI.

\bigskip
\section{Theoretical considerations}
\label{sec:1}

Let us consider a 5-dimensional theory with the action decomposed as \cite{Nihei}
\begin{equation}\label{01}
S= -\frac{1}{2\kappa^{2}_{(5)}}\int d^{5}x \sqrt{\bar{g}}[\bar{R}+2\Lambda_{(5)}] + \int d^{4}x \sqrt{-\bar{g}}L_{m}
\end{equation}
where, $\bar{g}_{AB}$ is a 5-dimensional metric of signature (+ - - - -). Here $\Lambda_{(5)}$ is the bulk cosmological constant and $\bar{R}$ is the 5-dimensional scalar curvature for the metric. The constant $\kappa_{(5)}$ is related to the 5-dimensional Newton's constant $G_{(5)}$ and the 5-dimensional reduced Planck mass $M_{(5)}$ by the relation
\begin{equation}\label{02}
\kappa^{2}_{(5)}= 8\pi G_{(5)}= M^{-3}_{(5)}.
\end{equation}

The 5-dimensional field equations for such a bulk are read as \cite{bdl1,bdl2,lrr}
\begin{equation}\label{04}
\bar{G}_{AB} = - \Lambda_{(5)}\bar{g}_{AB} + \kappa^{2}_{(5)}\bar{T}_{AB}
\end{equation}
where $\bar{G}_{AB}$ is the 5-dimensional Einstein tensor and $\bar{T}_{AB}$ represents the 5-dimensional energy-momentum tensor.

A general five-dimensional warped metric with flat spatial section and time-dependent warp factor can be written in the form
\begin{equation}\label{05}
dS^{2}=e^{2f(t,y)}\left(dt^2-R^2(t)(dr^2+r^2d\theta^2+r^2sin^{2}\theta d\phi^2)\right)-h(y)dy^2.
\end{equation}
where $y$ is the coordinate of the extra dimension, $t$ denotes the conformal time, $R(t)$ is the usual scale factor of the FRW metric and $h(y)$ is the scale factor of the extra dimension. We have considered a warped product space-time where the 4-dimensional hypersurface is a smooth four-manifold. Since the scale factor of the extra dimension depends only on the extra dimensional coordinate, it can always be rescaled to render the corresponding metric coefficient equal to unity. We assume that the fifth dimension is non-compact and curved (i.e. warped) \cite{rs2,lrr} and the warp factor is a function of both time, as well as of the extra dimensional coordinate \cite{GC}. Mathematically, the time dependence of the warp factor does not affect the smooth nature of the function $f$. Physically, it takes into account the possibility of a time-dependent process of localization of gravity. The effect of this time-dependence is pronounced during a particular phase of evolution. Thus the exponential warp factor $e^{2f(t,y)}$ reflects such a process of localization of gravity and may be related to the confining role of the time-varying bulk cosmological constant at such stages.

For the spacetime under consideration, the effective 4-dimensional metric has a scale factor $A$ which evolves as a composite function of conformal time, given by $A^{2}(\eta(t))=e^{2f(t,y=y_{0})}R^2(t)$, the observed universe being represented by the hypersurface $y=y_{0}$. Here, $\eta(t)\equiv\int dt e^{f(t,y=y_{0})}$ is the proper time of a co-moving observer, when the position of the hypersurface is fixed at $y=y_{0}$. The geometry of the observed universe at the location, $y=y_{0}$, will be determined by the metric

\begin{equation}\label{05a}
ds^{2}=g_{\alpha\beta}(x,y=y_{0})dx^{\alpha}dx^{\beta}=e^{2f_{0}}q_{\alpha \beta}(x)dx^{\alpha}dx^{\beta},
\end{equation}
where $f_{0}$ is the value of $f$ at $y=y_{0}$ and $q_{\alpha\beta}=q_{\alpha\beta}(x)$ is the warp metric on the 4-dimensional hypersurface.

As the hypersurface encounters "bending effect", the location of the hypersurface has some $y$-dependence. Geometrically, the extrinsic curvature of the hypersurface gives us a measure of the deviation of the hypersurface from the tangent plane and therefore such a bending may produce an observable result in the form of a smooth scalar function represented by the warp factor \cite{maia4}. Since this warp factor also depends on time, the deviation of the hypersurface from the tangent plane will change with time. The extrinsic curvature of the 4-dimensional hypersurface is defined by the relation
\begin{equation}\label{05b}
K_{\alpha\beta}=-\frac{1}{2}\pounds_{n}g_{\alpha\beta}(x^{\mu},y),
\end{equation}
where,  $x^{\mu}$ are the coordinates on the hypersurface and
\begin{equation}\label{05c}
K_{\alpha\beta;\gamma} - K_{\alpha\gamma;\beta}=0.
\end{equation}
For the metric of Eq. (\ref{05}) \cite{leon,dahia3,dahia4}, we have $K_{\alpha\beta} = -\frac{1}{2\sqrt{h}}\left(\frac{\partial g_{\alpha\beta}(x,y=y_{0})}{\partial y}\right)$. This condition guarantees the embedding of the four-dimensional manifold, so that, the extrinsic curvature $K_{\alpha\beta}$ also behaves as a physical field, as we shall see subsequently. This implies that the four-dimensional dynamical equations for the metric becomes affected by the presence of the extrinsic curvature.

In the present paper, we intend to study the effect of the time-dependent warp factor on the extrinsic curvature of the hypersurface and interpret the consequences from the physical point of view. In Section V, we find that the extrinsic curvature of the hypersurface is related to the vacuum energy-momentum tensor. The effective cosmological constant of the hypersurface depends on the curvature of the bulk metric, which is associated with the time-dependent process of bending of the hypersurface, due to the time-dependent nature of the extrinsic curvature. Thus the presence of the time-dependent warp factor effectively modifies the usual dynamics of the gravitational field compared to that predicted from Einstein's theory. If this modification takes place at today's Hubble scale, $H_{0}$, then it is expected that this will alter the gravitational force law at distance scales much smaller than $H_{0}^{-1}$ \cite{Arthur}, leading to an expansion history that can be identically reproduced by a dynamical dark energy model. It is evident that the gravitational force law is sensitive to the background cosmological expansion, since this expansion is intimately connected with the extrinsic curvature of the hypersurface \cite{Deff} and this curvature controls the effective Newton's constant.

It is known that if the flat 3-brane of the original RS setup is replaced by a dynamical brane, then the zero-mode graviton fluctuation is not necessarily localized on the brane, if the 5D bulk spacetime is anti de Sitter \cite{Neupane2}. That is, for an embedding of a $dS_4$ brane into a $AdS_5$ bulk, the massless graviton wave function is non-normalizable (even if it is a bound state solution) once the $Z_2$ symmetry is relaxed, so that the 4D Newton's constant is not finite. However, brane-world models with a positively curved $dS_5$ bulk generate a 4-dimensional cosmological constant in the gravity sector of the effective 4D theory with a finite 4D Newton's constant and satisfactorily explain the localization of the zero-mode graviton in 4 dimensions.

For the model under our consideration, the 4-dimensional effective action obtained under a dimensional reduction from 5D to 4D, is given by \cite{Lukas,Neupane2,Neupane3}
\begin{equation}
S_{eff} = - \frac{M^{3}_{(5)}}{2} \int_{\cal M}\sqrt{-g_4} \left[ R_4 + correction terms \right) = - \frac{M^{3}_{(5)}}{2} \int_{\cal M} \tau(t) \Gamma(y) \sqrt{h} \left( R_4 + ....\right).
\end{equation}
where $\cal M$ denotes the four-manifold. Thus, owing to the time-dependent nature of the warp factor, the 4D Newton's constant will be time-dependent. We have already pointed out that the process of localization of gravity becomes time-dependent. However, for the model to be physically realistic, the evolution of the universe should be such that it ultimately leads to the current state of the universe. Thus the time-dependent process of localization of gravity is expected to dominate only during the early phase of evolution of the universe.

To illustrate our study we choose to work with a specific example for the warp factor. For simplicity of calculations, we choose the scalar function $f(t,y)$ to be given by the simple form
\begin{equation}\label{06}
f(t,y)=\frac{1}{2}(\ln\tau(t)+\ln\Gamma(y)).
\end{equation}
With this choice, the five-dimensional metric assumes the form
\begin{equation}\label{07}
dS^{2}=\tau(t)\Gamma(y)\left(dt^2-R^2(t)(dr^2+r^2d\theta^2+r^2sin^{2}\theta d\phi^2)\right)-h(y)dy^2.
\end{equation}
This is a very specific choice of the bulk metric, for which the energy-momentum tensor has only the $T_{yy}$ component other than the components on the hypersurface. The components of the bulk stress-energy tensor are given by

\begin{center}
$\bar{T}^{t}_{t}=\bar{\rho},\qquad\qquad \bar{T}^{i}_{j}=-\bar{P},\qquad\qquad \bar{T}^{y}_{y}=-\bar{P}_{y}$.
\end{center}
where $\bar{\rho}$, $\bar{P}$ and $\bar{P}_{y}$ are the energy density and pressure in the bulk.

\section{Five-dimensional Field Equations}

To simplify the subsequent calculations, let us consider that the extra-dimensional coordinate has been rescaled to yield $h(y) \rightarrow 1$. The  non-vanishing components of the five-dimensional Einstein tensor for the warped product spacetime given by Eq.~(\ref{07}) (with this re-scaling), are obtained as

\begin{equation}\label{08}
\bar{G}_{tt}=\frac{3\bar{g}_{tt}}{4\tau\Gamma}\left( \frac{4 \dot{R}\dot{\tau}}{R \tau} + \frac{4 \dot{R}^2}{R^2} + \frac{\dot{\tau}^{2}}{\tau^{2}} \right) - \frac{6\bar{g}_{tt}}{4 }\frac{\Gamma^{\prime \prime}}{\Gamma},
\end{equation}

\begin{equation}\label{10}
\bar{G}_{yy}=\frac{3\bar{g}_{yy}}{4\tau\Gamma}\left( \frac{2\ddot{\tau}}{\tau} - \frac{\dot{\tau}^{2}}{\tau^{2}} + \frac{6 \dot{R} \dot{\tau}}{R \tau} + \frac{4 \dot{R}^2}{R^2} + \frac{4 \ddot{R}}{R} \right) - \frac{6\bar{g}_{yy}}{4 }\left( \frac{\Gamma^{\prime}}{\Gamma} \right)^{2},
\end{equation}

\begin{equation}\label{11}
\bar{G}_{ij}=\frac{\bar{g}_{ij}}{4\tau\Gamma} \left( \frac{4\ddot{\tau}}{\tau} - \frac{3\dot{\tau}^{2}}{\tau^{2}} + \frac{8 \dot{R} \dot{\tau}}{R \tau} + \frac{4 \dot{R}^2}{R^2} + \frac{8 \ddot{R}}{R} \right) - \frac{6\bar{g}_{ij}}{4 }\frac{\Gamma^{\prime \prime}}{\Gamma}.
\end{equation}

Above, an overdot represents derivative with respect to time $t$ and a prime stands for a derivative with respect to the fifth coordinate $y$. The five-dimensional field equations are obtained as

\begin{equation}\label{12}
\frac{3}{4\tau\Gamma}\left( \frac{\dot{\tau}^{2}}{\tau^{2}} + \frac{4 \dot{R}\dot{\tau}}{R \tau} + \frac{4 \dot{R}^2}{R^2} \right) - \frac{6}{4 }\frac{\Gamma^{\prime \prime}}{\Gamma} = - \Lambda_{(5)} + 8 \pi G_{(5)} \bar{\rho},
\end{equation}

\begin{equation}\label{13}
\frac{1}{4\tau\Gamma} \left( \frac{4\ddot{\tau}}{\tau} - \frac{3\dot{\tau}^{2}}{\tau^{2}} + \frac{8 \dot{R} \dot{\tau}}{R \tau} + \frac{4 \dot{R}^2}{R^2} + \frac{8 \ddot{R}}{R} \right) - \frac{6}{4 }\frac{\Gamma^{\prime \prime}}{\Gamma} = - \Lambda_{(5)} - 8 \pi G_{(5)} \bar{P},
\end{equation}

and

\begin{equation}\label{14}
\frac{3}{4\tau\Gamma}\left( \frac{2\ddot{\tau}}{\tau} - \frac{\dot{\tau}^{2}}{\tau^{2}} + \frac{6 \dot{R} \dot{\tau}}{R \tau} + \frac{4 \dot{R}^2}{R^2} + \frac{4 \ddot{R}}{R} \right) - \frac{6}{4}\left( \frac{\Gamma^{\prime}}{\Gamma} \right)^{2} = - \Lambda_{(5)} - 8 \pi G_{(5)} \bar{P}_{y}.
\end{equation}

In the calculations that now follow, we shall assume $8 \pi G_{(5)}=1$. From (\ref{12}) and (\ref{13}), we find that
\begin{equation}\label{14a}
\bar{\rho} + 3\bar{P} =  \frac{3}{4\tau\Gamma}\left( \frac{4 \dot{\tau}^{2}}{\tau^{2}} - \frac{4\ddot{\tau}}{\tau} - \frac{4 \dot{R} \dot{\tau}}{R \tau} - \frac{8 \ddot{R}}{R} \right) + \frac{12}{4}\left(  \frac{\Gamma^{\prime \prime}}{\Gamma} \right) - 2\Lambda_{(5)}
\end{equation}
and from (\ref{12})-(\ref{14}) we get
\begin{equation}\label{14b}
\bar{\rho} + 3\bar{P} + \bar{P}_{y} = \frac{3}{4\tau\Gamma}\left( \frac{5 \dot{\tau}^{2}}{\tau^{2}} - \frac{6 \ddot{\tau}}{\tau} - \frac{10 \dot{R} \dot{\tau}}{R \tau} - \frac{4 \dot{R}^2}{R^2} - \frac{12 \ddot{R}}{R} \right) + \frac{6}{4}\left( \frac{2\Gamma^{\prime \prime}}{\Gamma} + \left( \frac{\Gamma^{\prime}}{\Gamma} \right)^{2} \right) - 3\Lambda_{(5)}.
\end{equation}

From the above it is evident that the validity of the strong energy condition in the bulk will be governed by the nature of warping, as well as by the effect of the extra dimension.

\section{Solutions}

\subsection{Case 1: The case of vanishing Weyl tensor}

We find that the non-zero components of the Weyl tensor for this spacetime are of the form

\begin{equation}\label{15}
C_{ABAB}= constant \times \left(\frac{\bar{g}_{AA}\bar{g}_{BB}F(t)}{\tau\Gamma}\right),
\end{equation}
with
\begin{equation}\label{16}
F(t)= \left( \frac{2\ddot{\tau}}{\tau} - \frac{3 \dot{\tau}^{2}}{\tau^{2}} - \frac{2 \dot{R} \dot{\tau}}{R \tau} - \frac{4 \dot{R}^2}{R^2} + \frac{4 \ddot{R}}{R} \right).
\end{equation}
where, only the constant factor varies for the different components. Thus, if we have $F(t)=0$, the Weyl tensor for this space-time metric will vanish. Since we have only the equation (\ref{16}) to determine both $\tau(t)$ and $R(t)$, we have to impose additional constraints to satisfy this requirement. We presume that the effect of the time-dependent warp factor was pronounced during the early phase of evolution of the four-dimensional universe, namely the radiative phase, so that $R(t)=t^{1/2}$. Further, $\tau(t)$ must be a decreasing function of time, whose exact nature can be determined with the help of (\ref{16}). Let us consider a trial solution of the power-law type given by $\tau(t)=t^n$.  Substituting this in Eq. (\ref{16}) we obtain the relation

\begin{equation}\label{16a}
n^2 + 3n + 2 = 0,
\end{equation}
which yields $n= -1, -2$. We shall analyze the cosmological implication of such a solution in a subsequent section. However, the function $\Gamma$ cannot be found from this condition.

\subsection{Case 2: The case of a bulk with isotropic pressure}

Let us now consider a bulk with isotropic pressure, i. e. $\bar{P}=\bar{P}_{y}$. Using Eqs.~(\ref{13}) and (\ref{14}), we get the following condition satisfied by this bulk:

\begin{equation}\label{17}
\frac{1}{2 \tau} \left( \frac{\ddot{\tau}}{\tau}  + \frac{5 \dot{R} \dot{\tau}}{R \tau} + \frac{4 \dot{R}^2}{R^2} + \frac{2 \ddot{R}}{R} \right) + \frac{6 \Gamma}{4} \left( \frac{\Gamma^{\prime \prime}}{\Gamma} - \left( \frac{\Gamma^{\prime}}{\Gamma} \right)^{2} \right) = 0.
\end{equation}
Since $\tau$ is a function of time, whereas $\Gamma$ is a function of the extra dimensional coordinate, the Eq.~(\ref{17}) will be satisfied if the coefficient of $1/\tau$ and \textbf{$\Gamma$} vanish separately. The coefficient of $\Gamma$ will vanish when $\left( \frac{\Gamma^{\prime \prime}}{\Gamma} - \left( \frac{\Gamma^{\prime}}{\Gamma} \right)^{2} \right) = 0$. This leads us to the result

\begin{equation}\label{18}
\frac{1}{\Gamma} \frac{d \Gamma}{d y} = constant,
\end{equation}
which implies that
\begin{equation}\label{18b}
\Gamma(y) = C_{1}e^{\pm{y}},
\end{equation}
in conformity with known results. From Eq.~(\ref{17}), equating the coefficient of $1/\tau$ to zero, we get

\begin{equation}\label{19}
\left( \frac{\ddot{\tau}}{\tau}  + \frac{5 \dot{R} \dot{\tau}}{R \tau} + \frac{4 \dot{R}^2}{R^2} + \frac{2 \ddot{R}}{R} \right) = 0.
\end{equation}
Once again, considering a trial solution of the power-law type given by $\tau(t)=t^m$ and with $R(t)=t^{1/2}$, we obtain the relation

\begin{equation}\label{19a}
2m^2 + 3m + 1 = 0,
\end{equation}
which has the solution $m = -1/2, -1$.

Equations (\ref{18b}) and (\ref{19a}) give us particular solutions for the warp factor for this isotropic bulk. It appears that the solution $\tau(t) = t^{-1}$ will satisfy a bulk with isotropic pressure and vanishing Weyl tensor. However, it does not appear to be of much interest from the cosmological point of view, as we shall see later on.

Now, combining Eqs.~(\ref{12}) and (\ref{13}), we get

\begin{equation}\label{20}
\bar{\rho} + \bar{P} = \frac{1}{4\tau\Gamma} \left( - \frac{4\ddot{\tau}}{\tau} + \frac{6\dot{\tau}^{2}}{\tau^{2}} + \frac{4 \dot{R} \dot{\tau}}{R \tau} + \frac{8 \dot{R}^2}{R^2} - \frac{8 \ddot{R}}{R} \right).
\end{equation}
Considering the case of $\tau(t)=t^{-1/2}$ and $R=t^{1/2}$, we find that
\begin{equation}\label{21}
\bar{\rho} + \bar{P} = \frac{3}{8 \Gamma t^{3/2}}.
\end{equation}
From (\ref{21}), it is evident that the above sum will decrease as time increases, for a given value of $y$.

\section{Analysis of four-dimensional cosmology}

The effective four-dimensional metric is given by

\begin{equation}\label{23}
ds^2=\tau(t)\Gamma(y)\left(dt^2-R^2(t)(dr^2+r^2d\theta^2+r^2sin^{2}\theta d\phi^2)\right)=\tau(t)\Gamma(y)h_{\alpha \beta}(x)dx^{\alpha}dx^{\beta}.
\end{equation}
We find that the Weyl tensor for this metric is identically zero. The components of the extrinsic curvature for the hypersurface defined by Eq.~(\ref{23}) are obtained as
\begin{center}
$K_{tt}=-\frac{1}{2}\tau\Gamma^{\prime}$,
\end{center}
\begin{center}
$K_{rr}=\frac{1}{2}\tau\Gamma^{\prime} R^2(t)$,
\end{center}
\begin{center}
$K_{\theta\theta}=\frac{1}{2}\tau\Gamma^{\prime} R^2(t)r^{2}$
\end{center}
and
\begin{center}
$K_{\phi\phi}=\frac{1}{2}\tau\Gamma^{\prime} R^2(t)r^{2}\sin^{2}\theta$.
\end{center}
It is evident that the extrinsic curvature of this hypersurface is governed by the time-dependent warp factor. In compact form, we can write
\begin{equation}\label{24}
K_{\mu\nu}=-\frac{1}{2}\frac{\Gamma^{\prime}}{\Gamma}g_{\mu\nu}.
\end{equation}

Since $g_{\mu\nu}$ is diagonal, $K_{\mu\nu}$ also turns out to be diagonal. The time-dependent part of the warp factor is contained within $g_{\mu\nu}$. The equations of motion for this hypersurface embedded in a five-dimensional bulk, is derived directly from the Gauss and Codazzi equations for the conditions of integrability of the embedding geometry \cite{Eisenhart}. The result is basically the Einstein field equations, modified by the presence of the extra term $Q_{\mu\nu}$, which was obtained in \cite{maia1,maia2} after incorporating the junction conditions. The resulting field equations are quoted below:

\begin{equation}\label{25}
R_{\mu\nu}-\frac{1}{2}R g_{\mu\nu}+ \lambda g_{\mu\nu}= 8\pi G T_{\mu\nu} + Q_{\mu\nu}.
\end{equation}
We shall use this result for our subsequent analysis. The components of the 4-dimensional energy-momentum tensor are given by

\begin{center}
$T^{t}_{t}=\rho,\qquad\qquad and \qquad\qquad T^{i}_{j}=-p$.
\end{center}

For Eq.~(\ref{25}) to be valid, we must have
\begin{equation}\label{26}
Q^{\mu\nu}_{;\nu}=0.
\end{equation}
Together with the conservation of $T_{\mu\nu}$, this term $Q_{\mu\nu}$ turns out to be energetically uncoupled from the other components of the universe, whatever be its influence on the evolution of the hypersurface. The explicit form of this term is

\begin{equation}\label{27}
Q_{\mu\nu}=g^{\rho\sigma}K_{\mu\rho} K_{\nu\sigma}-\xi K_{\mu\nu} - \frac{1}{2}(K^{2}- \xi^{2})g_{\mu\nu}\\=K^{\sigma}_{\mu}K_{\nu\sigma}-\xi K_{\mu\nu} - \frac{1}{2}(K^{2}- \xi^{2})g_{\mu\nu},
\end{equation}
where $\xi=g^{\mu\nu}K_{\mu\nu}$ denotes the mean curvature and $K^{2}=K^{\mu\nu}K_{\mu\nu}$ is the Gaussian curvature of the hypersurface. This term $Q_{\mu\nu}$ does not exist in the Einstein's equations in pure Riemannian geometry \cite{maia2}. A quick look at (\ref{27}) reveals that it is indeed derived from the junction conditions. It is the contribution of the extrinsic curvature of the hypersurface, and is purely geometrical in origin.  Since $Q_{\mu\nu}$ also depends on $g_{\mu\nu}$, it is also affected by the time-dependent warp factor. Moreover, $Q_{\mu\nu}$ does not necessarily vanish, even when the bulk space-time is flat. Therefore, it effectively modifies the usual dynamics of the gravitational field compared to that predicted from Einstein's theory. The interaction between the bulk and the hypersurface occurs when a graviton crosses the hypersurface. It is subjected to a deviation, which is expressed in terms of the extrinsic curvature $K_{ij}$ of the embedded geometry, representing the tangential components of the local variation of the normal unit vector. Since the effective cosmological constant depends on the curvature of the bulk metric, \cite{maia1,maia2}, hence the dynamics of the space-time also responds to the vacuum energy. Thus, for the hypersurface under consideration, with the time-dependent nature of the extrinsic curvature, the vacuum energy is a time-dependent entity associated with the time-dependent process of bending of the hypersurface. Physically, this can be interpreted as the manifestation of a dynamical dark energy component.

In the remaining analysis, we assume that the sources on the hypersurface is in the form of a perfect fluid satisfying a linear equation of state. A straightforward calculation yields the result
\begin{equation}\label{28}
Q_{\mu\nu}= \frac{3}{4}\left( \frac{\Gamma^{\prime}}{\Gamma} \right)^2 g_{\mu\nu},
\end{equation}
which shows the dependence of $Q_{\mu\nu}$ on the warp factor through $g_{\mu\nu}$. The two non-trivial field equations are obtained as

\begin{equation}\label{29}
\frac{3}{4\tau\Gamma}\left( \frac{\dot{\tau}^{2}}{\tau^{2}} + \frac{4 \dot{R}\dot{\tau}}{R \tau} + \frac{4 \dot{R}^2}{R^2} \right) + \lambda = - 8\pi G \rho + \frac{3}{4}\left( \frac{\Gamma^{\prime}}{\Gamma} \right)^2
\end{equation}
and
\begin{equation}\label{30}
\frac{1}{4\tau\Gamma} \left( \frac{4\ddot{\tau}}{\tau} - \frac{3\dot{\tau}^{2}}{\tau^{2}} + \frac{8 \dot{R} \dot{\tau}}{R \tau} + \frac{4 \dot{R}^2}{R^2} + \frac{8 \ddot{R}}{R} \right) + \lambda = 8\pi G p + \frac{3}{4}\left( \frac{\Gamma^{\prime}}{\Gamma} \right)^2.
\end{equation}

\bigskip
The parameters $\rho$ and $p$ in the Eqs.~(\ref{29}) and (\ref{30}) are the total energy density and the total pressure of ordinary matter.
The effective energy density $\rho_{eff}$ of the matter on the hypersurface as obtained from Eq.~(\ref{29}) is given by
\begin{equation}\label{31}
\rho_{eff}=\rho + \lambda - \frac{3}{4}\left( \frac{\Gamma^{\prime}}{\Gamma} \right)^2 = \rho + \lambda_{eff},
\end{equation}
where, we have assumed that
\begin{center}
$8 \pi G = 1$ \quad \quad and \quad \quad $\lambda_{eff} = \lambda - \frac{3}{4}\left( \frac{\Gamma^{\prime}}{\Gamma} \right)^2 $.
\end{center}
Hence
\begin{equation}\label{31a}
\frac{3}{4\tau\Gamma}\left( \frac{\dot{\tau}^{2}}{\tau^{2}} + \frac{4 \dot{R}\dot{\tau}}{R \tau} + \frac{4 \dot{R}^2}{R^2} \right) = -\rho_{eff}.
\end{equation}

From Eq.~(\ref{30}), we get the effective pressure of this matter as
\begin{equation}\label{32}
p_{eff} = p - \lambda + \frac{3}{4}\left( \frac{\Gamma^{\prime}}{\Gamma} \right)^2 = p - \lambda_{eff},
\end{equation}
so that
\begin{equation}\label{32a}
\frac{1}{4\tau\Gamma} \left( \frac{4\ddot{\tau}}{\tau} - \frac{3\dot{\tau}^{2}}{\tau^{2}} + \frac{8 \dot{R} \dot{\tau}}{R \tau} + \frac{4 \dot{R}^2}{R^2} + \frac{8 \ddot{R}}{R} \right)= p_{eff}.
\end{equation}

Combining Eqs.~(\ref{31}) and (\ref{32}), we obtain
\begin{equation}\label{33}
\rho_{eff} + p_{eff} = \rho + p
\end{equation}
and
\begin{equation}\label{34}
\rho_{eff} + 3p_{eff} = \rho + 3p - 2\lambda_{eff}.
\end{equation}

From Eq.~(\ref{33}), it is evident that if ordinary matter obeys the weak energy condition (WEC), i.e.
\begin{center}
$\rho + p \geq 0 $,
\end{center}
then the matter in the four-dimensional part will also do the same, since in that case,
\begin{center}
$\rho_{eff} + p_{eff} \geq 0$.
\end{center}
Hence, both may represent a quintessence field. Further, to have $\rho_{eff} > 0$, we must have
\begin{equation}\label{34b}
\rho + \lambda_{eff} > 0,
\end{equation}
thereby setting a restriction on the value of $\lambda_{eff}$. For physically possible matter fields, $\rho > 0$, but unless $\rho > |\lambda_{eff}|$, we cannot have $\rho_{eff} > 0$. Moreover, even when $\rho_{eff} > 0$, we may have $p_{eff}<0$. In that case, assuming that the ordinary matter is represented by an equation of state of the form $p=w\rho$, we obtain from (\ref{32}),
\begin{equation}\label{34c}
w\rho - \lambda_{eff} < 0.
\end{equation}

From Eqs.~(\ref{34b}) and (\ref{34c}), we obtain the following restriction on the value of $\rho$:

\begin{equation}\label{34d}
|\lambda_{eff}|< \rho < |\frac{\lambda_{eff}}{w}|,
\end{equation}
for which the effective matter on the hypersurface will experience negative pressure. Therefore, we conclude that, unless we have the ordinary matter as a phantom field, we cannot have the effective matter in the four-dimensional part to behave as a phantom field, i.e. the behavior of the effective matter on the hypersurface follows the behavior of the ordinary matter in the higher dimensional scenario in this case. The effective matter will be described by a negative pressure quintessence field provided $\rho$ satisfies the condition in Eq.~(\ref{34d}).

For the case of dust, we obtain
\begin{equation}
\lambda_{eff}= \frac{1}{4\tau\Gamma} \left( \frac{3\dot{\tau}^{2}}{\tau^{2}} - \frac{4\ddot{\tau}}{\tau} - \frac{8 \dot{R} \dot{\tau}}{R \tau} - \frac{4 \dot{R}^2}{R^2} - \frac{8 \ddot{R}}{R} \right),
\end{equation}
which shows that the effective cosmological constant on the hypersurface is a variable quantity, monitored by the warp factor and can be interpreted as a dynamical dark energy component.

Thus the usual four-dimensional Friedmann equation, relating the expansion rate $H$ of our universe with the energy density of the universe, will get modified by the presence of the extra geometric term arising from the effect of warping. In spite of that, we find that the evolution closely follows the standard predictions. From the expression of the scale factor of this four-dimensional universe, we obtain the Hubble parameter for the effective metric as
\begin{equation}\label{34a}
H = \left( \frac{\dot{R}}{R} + \frac{\dot{\tau}}{2 \tau} \right).
\end{equation}
Thus we have
\begin{equation}\label{34a1}
\rho = \frac{3H^{2}}{\tau\Gamma} + \frac{3}{4}\left( \frac{\Gamma^{\prime}}{\Gamma} \right)^{2} - \lambda.
\end{equation}
The last equation is very significant in the sense that when the warp factor is reduced to a constant, it resembles the usual results of standard cosmology. From Eq.~(\ref{34a1}) we get
\begin{equation}\label{34a2}
\frac{\rho}{3H^2} + \frac{\lambda}{3H^2} - \frac{3}{12H^2}\left( \frac{\Gamma^{\prime}}{\Gamma} \right)^{2} = \frac{1}{\tau\Gamma},
\end{equation}
which is indeed the equation involving the density parameters of the various components in the four-dimensional universe. It is evident that the standard equation gets modified by the effect of the warp factor.

The explicit form of the deceleration parameter is now
\begin{equation}\label{35}
q = -1 - \frac{\dot{H}}{H^{2}} = -1 - \frac{2 R \ddot{R} \tau^2 + R^2 \tau \ddot{\tau} - 2 \dot{R}^2 \tau^2 - R^2 \dot{\tau}^2 }{(2 \dot{R} \tau + R \dot{\tau} )^2}.
\end{equation}

From the above analysis, we can draw the following conclusions:
\begin{itemize}
\item \textbf{Status of energy conditions}: From Eq.~(\ref{34}) it is evident that although ordinary matter in the higher-dimensional scenario may obey the strong energy condition (SEC), the matter in the effective four-dimensional part may not do so, unless appropriate conditions are satisfied by $\lambda_{eff}$. Hence for some value of $\lambda$ and $\Gamma$, the effective matter may violate the SEC and behave as dark energy, although ordinary matter in the higher dimensional scenario may not behave as such. Such a thing will happen if $\rho + 3p \geq 0$, but $\rho + 3p < 2\lambda_{eff}$ i.e., $\rho + 3p < 2 \left[\lambda + \frac{3}{4}\left( \frac{\Gamma^{\prime}}{\Gamma} \right)^2 \right]$, so that $\rho_{eff} + 3p_{eff} < 0$. Since $\lambda$ is a variable quantity, monitored by the warp factor, the status of the energy conditions also depend on the effect of the warp factor. It must be noted here, that this strange behavior of the effective matter is basically a consequence of the embedding scheme.

\item \textbf{Deceleration parameter}:We know that for the Friedmann models, the two field equations together produce the result
\begin{center}
$\frac{\ddot{R}}{R}=\frac{\lambda}{3}-\frac{1}{6}(\rho+3p)$.
\end{center}
The deceleration parameter is given by
\begin{center}
$q= - \frac{\ddot{R}}{RH^{2}}$.
\end{center}
Therefore,
\begin{center}
$q=\frac{1}{6H^{2}}(\rho + 3p) - \frac{\lambda}{3H^{2}}$.
\end{center}
This means that, even if $\rho + 3p >0$, $q$ may not be positive because that will also depend on the contribution of the cosmological term, and accordingly there may be either acceleration or deceleration. However, if $\rho + 3p < 0$, then $q<0$ and the universe will exhibit acceleration. Moreover, the effective matter will undergo acceleration even when the ordinary matter in the higher-dimensional scenario may not do so.
\end{itemize}

\bigskip
\textbf{Specific cases:} Let us now examine the two solutions in the case of isotropic bulk, viz.
\begin{enumerate}
\item $\tau(t)=t^{-1}$ and $R(t)=t^{1/2}$
\item $\tau(t)=t^{-1/2}$ and $R(t)=t^{1/2}$
\end{enumerate}

\emph{Case 1:} In this case, the universe is found to have a constant value for the scale factor $A$ and zero value for the Hubble parameter $H$ and is therefore ruled out from physical consideration.

\emph{Case 2:} Here, the hubble parameter decreases with time, being inversely proportional to $t$. The deceleration parameter is found to be $ q = - 1 + 1/3t $. It is positive at very small values of time but changes sign at some epoch of time, and finally settles to a value of -1 after sufficiently large time. Thus the universe is initially decelerated, but subsequently makes a transition to an accelerated phase at later times. The effective energy density and effective pressure are given by
\begin{equation}\label{36a}
\rho_{eff} = \frac{3}{16 \Gamma t^{3/2}}
\end{equation}
and
\begin{equation}\label{36b}
p_{eff} = -\frac{3}{16 \Gamma t^{3/2}},
\end{equation}
indicating that the effective matter violates the strong energy condition, although it still satisfies the weak energy condition. The universe will be dominated by vacuum energy after sufficiently long time.

\section{Summary and conclusions}

This paper deals with a five-dimensional warped product space-time having time-dependent warp factor and a non-compact fifth dimension. We know that for the RS models, the warp factor reflects the confining role of the bulk cosmological constant to localize gravity at the hypersurface through the curvature of the bulk. This process of localization may include some time-dependence during a particular stage of evolution of our universe. Hence we have considered a warp factor which depends both on time as well as on the extra coordinate. To simplify the derivation, we have assumed the warp factor in the product form: a function of time and a function of the extra-dimensional coordinate $y$. The 4-dimensional hypersurface representing the ordinary universe is defined by a flat FRW-type metric in the spatial dimension. Subsequently the solutions for the five-dimensional bulk have been determined for the case of vanishing Weyl tensor and separately when the pressure in the bulk is isotropic. When the scale of this extra dimension is properly fixed, the y-dependent part of the warp factor is simply an exponential function of the extra dimensional coordinate, in conformity with known results.

For the effective four-dimensional metric, the extrinsic curvature is found to be governed by the time-dependent warp factor. Following the method of Maia \cite{maia1,maia2}, we have computed the field equations for the effective matter with one extra term in the matter component, which effectively modifies the usual dynamics of the gravitational field compared to that predicted from Einstein's theory. This extra matter term depends on the geometric quantities, namely the extrinsic curvature, the mean curvature and the metric coefficients. Thus the effective cosmological constant of the 4-dimensional hypersurface is a variable quantity, monitored by the warp factor, leading to a geometric interpretation of a dynamical dark energy. The gravitational force law is sensitive to the background cosmological expansion, since this expansion is intimately connected with the time-dependent extrinsic curvature of the hypersurface, which in turn controls the effective 4D Newton's constant. The zero-mode graviton fluctuation is no longer guaranteed to be localized on the brane and the 4D Newton's constant is not finite. Thus, owing to the time-dependent nature of the warp factor, the process of localization of gravity becomes time-dependent and the 4D Newton's constant will also be time-dependent. This is the most significant consequence of considering a time-dependent process of localization of gravity. It automatically generates a dynamical dark energy component in the effective four-dimensional universe, without the necessity of incorporating anything by hand. Although the standard equation for the density parameter is modified by the effect of the warp factor, in spite of that, the evolution closely follows the standard predictions. We find that if the ordinary matter satisfies (or violates) the weak energy condition, then the effective matter also does so, but in the case of the strong energy condition, the situation may not be identical. Depending on the warp factor, the effective matter may violate the strong energy condition but the ordinary matter may still obey it. For a specific solution corresponding to an isotropic bulk, the universe is initially decelerated, but subsequently makes a transition to an accelerated phase at later times, being dominated by vacuum energy after sufficiently long time.

We conclude by stating that here we have illustrated a geometrical interpretation of dark energy in the universe emerging from a simple type of time-dependent warping in a warped product bulk. For future work we are considering the effect of a general type of time-dependent warp factor.

\section*{Acknowledgments}
SG thanks IUCAA, India for an associateship. SC is thankful to CSIR, Govt. of India for funding a project (03(1131)/08/EMR-II). Some part of these calculations were done with the help of the GRTensor package \cite{lake}. We are thankful to Prof. K. Lake and Prof. A. Banerjee for their valuable comments during the initial part of this work. We are also thankful to the reviewers for their valuable comments and suggestions.


\begin{thebibliography}{00}
\bibitem{kk} T. Kaluza, {\it Zum Unit\"{a}tsproblem der Physik}, Sitz. Preuss. Akad. Wiss. Phys. Math. {\bf K1} 966 (1921).
\bibitem{rbvspv} V. A. Rubakov and M. E. Shaposhnikov, {\it Phys. Lett.} {\bf 125B} 139 (1983).
\bibitem{aadd} N. Arkani-Hamed, S. Dimopoulos and G. Dvali, {\it Phys. Lett B} {\bf 429} 263 (1998); I. Antoniadis, N. Arkani-Hamed, S. Dimopoulos, G. Dvali, {\it Phys. Lett B} {\bf 436} 257 (1998).
\bibitem{add} N. Arkani-Hamed, S. Dimopoulos and G. Dvali, {\it Phys. Rev. D} {\bf 59} 086004 (1999).
\bibitem{khoury} J. Khoury, B. A. Ovrut, P. J. Steinhardt, and N. Turok, {\it Phys. Rev. D} {\bf 64} 123522 (2001).
\bibitem{noncompact} D. W. Joseph,  {\it Phys. Rev.} {\bf 126}, 319 (1962); K. Akama, {\it Lect. Notes Phys.} {\bf 176}, 267 (1982); M. Visser, {\it Phys. Lett. B} {\bf 159}, 22 (1985); E. J. Squires, {\it Phys. Lett. B} {\bf 167}, 286 (1986); G. W. Gibbons and D. L. Wiltshire, {\it Nucl. Phys. B} {\bf 287} 717 (1987); P. S. Wesson, {\it Ap. J.} {\bf 394} 19 (1992); P. S. Wesson and J. Ponce de Leon, {\it J. Math. Phys.} {\bf 33}, 3883 (1992); A. Banerjee, S. B. Dutta Choudhury and S. Chatterjee, {\it Gen. Rel. Grav.} {\bf 24} 991 (1992); A. A. Coley, {\it Ap. J.} {\bf 427} 585 (1994).
\bibitem{Gogberashvili} M. Gogberashvili, {\it Int. J. Mod. Phys. D} {\bf 11} 1635 (2002); Europhys. Lett. {\bf 49} 396 (2000); {\it Mod. Phys. Lett. A} {\bf 14} 2025 (1999); {\it Int. J. Mod. Phys. D} {\bf 11} 1639 (2002).
\bibitem{others} I. Antoniadis, {\it Phys. Lett. B} {\bf 246}, 377 (1990); J.D. Lykken, {\it Phys. Rev. D} {\bf 54}, 3693 (1996); R. Sundrum, {\it Phys. Rev. D} {\bf 59}, 085009 (1999) to name a few.
\bibitem{HW} P. Horava and E. Witten, {\it Nucl. Phys. B} {\bf 475}, 94 (1996).
\bibitem{Lukas} A. Lukas, B. A. Ovrut and D. Waldram, {\it Phys. Rev. D} {\bf 60} 086001 (1999); \emph{ibid.} {\bf 61} 023506 (2000).
\bibitem{cosmo1} N. Arkani-Hamed, S. Dimopoulos and J. March-Russell, {\it Phys. Rev. D} {\bf 63} 064020 (2001); N. Arkani-Hamed, S. Dimopoulos, N. Kaloper and J. March-Russell, {\it Nucl. Phys. B} {\bf 567} 189 (2000); C. Csaki, M. Graesser and J. Terning, {\it Phys. Lett. B} {\bf 456} 16 (1999); E. E. Flanagan, S. -H. H. Tye and I. Wasserman, {\it Phys. Rev. D} {\bf 62} 024011 (2000).
\bibitem{cosmo2} G. Dvali and S. H. Tye, {\it Phys. Lett. B} {\bf 450}, 72 (1999); G. Dvali and G. Gabadadze, {\it Phys. Lett. B} {\bf 460}, 47 (1999); G. Dvali and M. Shifman, {\it Phys. Rep} {\bf 320} 107 (1999); T. Banks, M. Dine and A. Nelson, {\it J. High Energy Phys.} {\bf 06}, 014 (1999).
\bibitem{other1} D. H. Lyth, {\it Phys. Lett. B} {\bf 448}, 191 (1999); N. Kaloper and A. Linde, {\it Phys. Rev. D} {\bf 59}, 101303 (1999); J. M. Cline, \emph{ibid.} {\bf 61} 023513 (2000).
\bibitem{other2} H. A. Chamblin and H. S. Reall, {\it Nucl. Phys. B} {\bf 562} 133 (1999); N. Kaloper, {\it Phys. Rev. D} {\bf 60}123506 (1999); P. Kanti and K. A. Olive, {\it Phys. Rev. D} {\bf 60} 043502 (1999); and many others.
\bibitem{rs1} L. Randall and R. Sundrum, {\it Phys. Rev. Lett.} {\bf 83} 3370 (1999).
\bibitem{rs2} L. Randall and R. Sundrum, {\it Phys. Rev. Lett.} {\bf 83} 4690 (1999).
\bibitem{LR} J. Lykken and L. Randall, {\it J. High Energy Phys.} {\bf 06} 014 (2000).
\bibitem{ADDK} N. Arkani-Hamed, S. Dimopoulos, G. Dvali and N. Kaloper, {\it Phys. Rev. Lett.} {\bf 84} 586 (2000); C. Cs\'aki and Y. Shirman, {\it Phys. Rev. D} {\bf 61} 024008 (2000); I. Oda, {\it Phys. Lett. B} {\bf 480} 305 (2000); A. Chodos and E. Poppitz,  \emph{ibid.} {\bf 471} 119 (1999).
\bibitem{otherRS} P. Kraus, {\it J. High Energy Phys.} {\bf 12}, 011 (1999); J. Cline, C. Grojean and G. Servant, {\it Phys. Lett. B} {\bf 472} 302 (2000); C. Grojean, J. Cline, and G. Servant, {\it Nucl. Phys. B} {\bf 578} 259 (2000); I. Oda, {\it Phys. Lett. B} {\bf 472} 59 (2000); T. Li, \emph{ibid.} {\bf 471} 20 (1999).
\bibitem{bdl1} P. Binetruy, C. Deffayet and D. Langlois,  {\it Nucl. Phys. B} {\bf 565} 269 (2000).
\bibitem{bdl2} P. Binetruy, C. Deffayet, U. Ellwanger and D. Langlois, {\it Phys. Lett. B} {\bf 477} 285 (2000).
\bibitem{Nihei} T. Nihei, {\it Phys. Lett. B} {\bf 465} 81 (1999).
\bibitem{GW} W.D. Goldberger and M.B. Wise, {\it Phys. Rev. D} {\bf 60} 107505 (1999).
\bibitem{LS} M.A. Luty and R. Sundrum, Phys. Rev. D {\bf 62} 035008 (2000).
\bibitem{KKOP} P. Kanti, I. I. Kogan, K. A. Olive, and M. Pospelov, {\it Phys. Lett. B} {\bf 468} 31 (1999).
\bibitem{CGRT} C. Cs\'aki, M. Graesser, L. Randall, and J. Terning, {\it Phys. Rev. D} {\bf 62} 045015 (2000).
\bibitem{CGKT} C. Cs\'aki, M. Graesser, C. Kolda, and J. Terning, {\it Phys. Lett. B} {\bf 462} 34 (1999).
\bibitem{CGS} J.M. Cline, C. Grojean, and G. Servant, {\it Phys. Rev. Lett.} {\bf 83} 4245 (1999).
\bibitem{Neupane1} I. P. Neupane, {\it Int. Jour. Mod. Phys. D} {\bf 19} 2281 (2010).
\bibitem{lrr} R. Maartens {\it Liv. Rev. Rel.} 2004-7 (http://www.livingreviews.org/lrr-2004-7); R. Maartens and K. Koyama [arXiv:1004.3962 [hep-th]].
\bibitem{GC} Such a warp factor has been considered by the authors in a recent paper. See Guha S and Chakraborty S {\it Gen. Relativ. Grav.} {\bf 42} 1739 (2010).
\bibitem{maia4} M. D. Maia and G. S. Silva, {\it Phys. Rev. D} {\bf 50} 7233 (1994).
\bibitem{leon} J. P. Leon, {\it Gen. Rel. Grav.} {\bf 36} 923 (2004).
\bibitem{dahia3} F. Dahia, C. Romero, L. F. P. Silva and R. Tavakol, {\it J. Math. Phys.} {\bf 48} 072501 (2007).
\bibitem{dahia4}  F. Dahia, C. Romero, L. F. P. Silva and R. Tavakol, {\it Gen. Rel. Grav.} {\bf 40} 1341 (2008).
\bibitem{Arthur} A. Lue, R. Scoccimarro and G. D. Starkmann, {\it Phys. Rev. D} {\bf 69} 124015 (2004).
\bibitem{Deff} C. Deffayet, {\it Phys. Lett. B} {\bf 502} 199 (2001); G. Dvali, G. Gabadadze, and M. Porrati, {\it Phys. Lett. B} {\bf 485}
208 (2000).
\bibitem{Neupane2} I. P. Neupane, {\it Phys. Rev. D} {\bf 83} 086004 (2011).
\bibitem{Neupane3} I. P. Neupane, {\it Class. Quant. Grav.} {\bf 26} 195008 (2009); {\it Phys. Lett. B} {\bf 683} 88 (2010).
\bibitem{Eisenhart} L. P. Eisenhart, Riemannian Geometry (Princeton University Press, Princeton (1949) pp. 149).
\bibitem{maia1} M. D. Maia, E.M. Monte and J. M. F. Maia, {\it Phys. Lett. B} {\bf 585} 11 (2004). The junction conditions have been incorporated by these authors to arrive at equation (\ref{25}) in this paper and also in the next paper.
\bibitem{maia2} M. D. Maia, E.M. Monte, J. M. F. Maia and J. S. Alcaniz, {\it Class. Quant. Grav.} {\bf 22} 1623 (2005).
\bibitem{lake}K. Lake and P. J. Musgrave, GRTensor (Queen's University, Kingston, 2003).
\end{thebibliography}
\end{document}